\documentclass{ws-procs975x65}

\begin{document}

\title{Non-axisymmetric configurations in the post-Newtonian Approximation to General Relativity }

\author{Norman G{\"u}rlebeck}

\address{Institute of Theoretical Physics, Charles University,\\
Prague, Czech Republic}

\author{David Petroff}

\address{Institute of Theoretical Physics, Friedrich Schiller University,\\
Jena, Germany}

\begin{abstract}
We use surface adapted ellipsoidal coordinates to obtain a first order post-Newtonian (PN) approximation to Dedekind ellipsoids with the intention of proceeding to higher orders.
\end{abstract}


\bodymatter

\section{Introduction}

Two effects may prevent the existence of non-axisymmetric, but nonetheless stationary configurations in General Relativity. In the case of imperfect fluids, dissipation will lead to a heating of the configuration and thus to time-dependent behavior. This was rigorously proved in Ref.~\citen{Lindblom_1976}. However, dissipation is completely neglected in perfect fluids, but the configuration can still lose energy due to gravitational radiation. The energy loss is known from the linearized theory to be associated with the time variation of the Newtonian quadrupole moment. By basing one's considerations on a Newtonian mass distribution which is non-axisymmetric, 
but has a constant quadrupole moment, one might hope to find a fully relativistic solution, which is also non-axisymmetric but does not contain gravitational radiation. The possibility or impossibility of such configurations has not been proved yet. 

Since solving the complete problem is not feasible, a PN approximation of an appropriate Newtonian solution, the Dedekind ellipsoids, is performed. 
These ellipsoids are triaxial with the axes $a_1>a_2>a_3$; the matter content is described by a perfect fluid, whose elements move on ellipses perpendicular to the shortest axis. The sequence of Dedekind ellipsoids forms a one parameter family, where for example $\frac{a_2}{a_1}\in(0,1)$ serves as the parameter. Chandrasekhar and Elbert investigated the deformed figures of the Dedekind ellipsoids in the PN approximation to first order \cite{Chandra_1974,Chandra_1978}. These calculations were carried out in Cartesian coordinates relying heavily on Newtonian results obtained in Ref.~\citen{Chandra_EFE}. This approach would become increasingly cumbersome for higher order corrections. On the other hand, the PN approximation of Maclaurin spheroids \cite{Petroff_2003} shows that spheroidal coordinates can be implemented advantageously. Therefore, we attempt to do the same here and employ ellipsoidal coordinates, at least in a surface adapted form. 

The connection to known solutions can be made by requiring that in the limit of axial symmetry, $\frac{a_2}{a_1}\to 1$, the PN approximation of the Dedekind ellipsoids be equal to the PN approximation of the Maclaurin spheroids at the bifurcation point ($1.716...\, a_3=a_1=a_2$). This limit is thoroughly investigated in Ref.~\citen{Petroff_2010} for the post-Newtonian Dedekind ellipsoids obtained in Ref.~\citen{Chandra_1978}. The other end of the Dedekind sequence ($\frac{a_2}{a_1}\to 0$) describes a static line of mass; therefore, it generates a Weyl field.

\section{The formalism and the equations}

To exploit the stationarity maximally, we use the projection formalism developed by Geroch in Ref.~\citen{Geroch_1971}. If the time coordinate $x^{0}$ is chosen to be Killing, the metric is decomposed into $F=g_{00}$, $h_{ab}=g_{ab}$ and $\xi_a=g_{0a}$, where $\xi_{\mu}$ is the Killing vector and $g_{\mu\nu}$ the metric. Greek indices run from $0$ to $3$ and Latin from $1$ to $3$.  Furthermore, we have a mass density $\rho$, which we assume to be a constant not subject to the PN approximation, a four-velocity $u^{\mu}$ and a pressure $p$. These functions are expanded in some dimensionless, relativistic parameter $\epsilon\ll 1$.  The PN expansion of a quantity $f$ is then given by $f=\sum\limits_{n=n_0}^\infty f^{(n)}\epsilon^{n}$, where $f^{(n_0)}$ denotes the Newtonian quantity. 

We use surface adapted ellipsoidal coordinates. The common ellipsoidal coordinates $\{\lambda,\mu,\nu\}$, as defined e.g.\ in Ref.~\citen{Byerly}, have the following coordinate surfaces: $\lambda=\mathrm{constant}$ describes confocal ellipsoids, $\mu=\mathrm{constant}$ confocal one-sheeted hyperboloids and $\nu=\mathrm{constant}$ confocal two-sheeted hyperboloids. The surface of a Newtonian Dedekind ellipsoid is given by $\lambda=a_1$. Assuming we were to have already solved the entire problem, we can write the surface of the resulting configuration as  $\lambda(\mu,\nu)=a_1 (1+\sum\limits_{n=2}^{\infty} S^{(2n)}(\mu,\nu)\epsilon^{2n})$. By introducing a new coordinate 
\begin{align}
 \lambda'= \lambda  (1+\sum\limits_{n=2}^{\infty} S^{(2n)}(\mu,\nu)\epsilon^{2n})^{-1},
\end{align}
the surface of the configuration is still defined by $\lambda'=a_1$.

In these coordinates, $h^{(0)}_{ab}$ is the Euclidean metric in ellipsoidal coordinates and the next non-vanishing order $h^{(2)}_{ab}$ is determined by a linear system of second order partial differential equations, whose solution is given by $-2 F^{(2)} h^{(0)}_{ab}$, where $ F^{(2)} $ denotes the Newtonian potential. The equations for $\xi^{(3)}_a$ can in an appropriate gauge be reduced to a Poisson equation where the inhomogeneity is given by the Newtonian velocity distribution. This equation can be solved by separation of variables and an expansion in ellipsoidal surface harmonics \cite{Byerly}. Therefore, $\xi^{(3)}_a$ and $h^{(2)}_{ab}$ are obtained without using the PN corrections of the surface, the four-velocity and the pressure.

The structure of the equation for $F^{(4)}$ is given by
\begin{align}\label{eq:F4}
 \Delta F^{(4)}=g_1(S^{(2)},f^{(n_0)},x^{a}),
\end{align}
where the function $g_1$ depends on Newtonian quantities $f^{(n_0)}$, on the coordinates $x^{a}$ and linearly on $S^{(2)}$; $\Delta$ denotes the Laplace operator in Euclidean space. If $S^{(2)}$ is expanded in ellipsoidal surface harmonics, Eq. \eqref{eq:F4} can be solved and the solution is determined by the expansion coefficients of $S^{(2)}$. The boundary conditions for Eq. \eqref{eq:F4} are given by the assumption of asymptotic flatness. The functions describing the matter and $F^{(4)}$ can be further restricted using the Bianchi identity
\begin{subequations}
\begin{align}
 u^{(2)a}_{\phantom{(2)a},a} &=g_{2}(f^{(n_0)},x^{a})\label{eq:Bianchidiv},\\
\nabla p^{(2)}&=g_{3}(F^{(4)},S^{(2)},u^{(2)},f^{(n_0)},x^{a})\label{eq:Bianchigrad}, 
\end{align}
\end{subequations}
where $g_3$ is again linear in the PN quantities. These equations have to be considered in the matter configuration, i.e.\ in the Newtonian ellipsoids, because of our surface adapted coordinates. 
The integrability condition of Eq. \eqref{eq:Bianchigrad} yields a equation relating $S^{(2)}$ and $u^{(2)}$ provided that  the general solution for $F^{(4)}$ is inserted. Furthermore, Eq. \eqref{eq:Bianchigrad} determines the PN correction of the pressure up to a constant. Eq. \eqref{eq:Bianchidiv} restricts the four-velocity, but does not determine it completely. There is still the freedom to add a source-free vector field to the spatial components of the four-velocity. There are two more conditions, which are satisfied by a stationary solution. The pressure vanishes at the surface as well as the component of the four-velocity normal to the surface. These conditions take on a particularly simple form in the coordinate system mentioned above since they reduce to $p(\lambda'=a_1,\mu,\nu)=0$ and $u^{\lambda'}(\lambda'=a_1,\mu,\nu)=0$. Solving these equations to the first PN order, the four-velocity and the surface corrections are further fixed.

\section*{Acknowledgements}

We gratefully acknowledge helpful discussions with M.\ Ansorg, J.\ Bi\v c\'ak, J.\ Friedman and R.\ Meinel.
The first author was financially supported by the grants GAUK 116-10/258025 and GACR 205/09/H033 and the second
by the Deutsche Forschungsgemeinschaft as part of the project ``Gravitational Wave Astronomy'' (SFB/TR7--B1).


\begin{thebibliography}{99}
 \bibitem{Byerly} W.E. Byerly, {\em Treatise on Fourier Series and Spherical, Cylindrical and Ellipsoidal Harmonics}, Dover publ., New York, 1893
	\bibitem{Chandra_EFE} S. Chandrasekhar, {\em Ellipsoidal Figures of Equilibrium}, (Dover publ., New York, 1987)
	
	\bibitem{Chandra_1974} S. Chandrasekhar and D. Elbert,
 {\em Astrophysical Journal} {\bf 192},731--746 (1974)
     
	\bibitem{Chandra_1978} S. Chandrasekhar and D. Elbert,
 {\em Astrophysical Journal} {\bf 220}, 303--313 (1978)

\bibitem{Geroch_1971} R. Geroch, {\em J.Math.Phys.} {\bf 12}, 1918--1924 (1971)

\bibitem{Lindblom_1976} L. Lindblom, {\em Astrophysical Journal} {\bf 208}, 873--880 (1976)

\bibitem{Petroff_2003} D. Petroff, {\em Phys. Rev. D} {\bf 68}, 104029 (2003)

\bibitem{Petroff_2010} D. Petroff and N. G{\"u}rlebeck, in preparation

\bibitem{Stephani} H. Stephani et. al., {\em Exact Solutions of Einstein's Field Equations}, (Cambridge University Press, Cambridge, 2003)



\end{thebibliography}
\end{document}